\documentclass[12pt]{iopart}
\usepackage{graphicx}%
\usepackage{listings}%
\usepackage{multirow}%

\expandafter\let\csname equation*\endcsname\relax
\expandafter\let\csname endequation*\endcsname\relax
\usepackage{amsmath}
\usepackage{xspace}
\usepackage{mcite}
\usepackage{subcaption}%
\usepackage[font=small]{caption}

\usepackage[ruled]{algorithm2e} 

\SetAlFnt{\small}
\SetAlCapFnt{\small}
\SetAlCapNameFnt{\small}
\SetAlCapHSkip{0pt}
\IncMargin{-\parindent}

\begin{document}

\title[Context-Aware Coupler Reconfiguration for Tunable Coupler-Based SQC]{Context-Aware Coupler Reconfiguration for Tunable Coupler-Based Superconducting Quantum Computers}

\author{Leanghok Hour, Sovanmonynuth Heng, Sengthai Heng, Myeongseong Go, Youngsun Han}

\address{Department of AI Convergence, Pukyong National University, Busan, South Korea}
\ead{youngsun@pknu.ac.kr}
\vspace{10pt}

\begin{indented}
\item[]March 2024
\end{indented}

\begin{abstract}
We address interconnection challenges in limited-qubit superconducting quantum computers (SQC), which often face crosstalk errors due to expanded qubit interactions during operations. 
Existing mitigation methods carry trade-offs, like hardware couplers or software-based gate scheduling.
Our innovation, the Context-Aware COupler REconfiguration (CA-CORE) compilation method, aligns with application-specific design principles. 
It optimizes the qubit connections for improved SQC performance, leveraging tunable couplers. 
Through contextual analysis of qubit correlations, we configure an efficient coupling map considering SQC constraints.
Our method reduces depth and SWAP operations by up to 18.84\% and 42.47\%, respectively. 
It also enhances circuit fidelity by 40\% compared to IBM and Google's topologies. 
Notably, our method compiles a 33-qubit circuit in less than 1 second.
\end{abstract}

%
\noindent{\it Keywords}: quantum computing, domain-specific application, computer architecture
%
%
%
%

\section{Introduction}
\label{sec:introduction}

A new computing paradigm of quantum computing leverages the features of quantum mechanics like superposition and entanglement with the potential to solve classically intractable problems much faster, such as unstructured search \cite{grover1996af} and integer factorization \cite{shor1994af}.
Currently, we are in the era of noisy intermediate-scale quantum (NISQ) devices with various hardware technologies namely superconducting~\cite{mooney2021wd} and trapped-ion~\cite{murali2020an} systems. 
Quantum computing vendors have all announced their quantum devices, including IBM's Osprey (433 qubits)~\cite{choi2023iq}, Google's Sycamore (53 qubits)~\cite{arute2019qs}, Rigetti's Aspen-M-3 (79 qubits)~\cite{gustafson2023pq}, and IonQ's Forte (32 qubits)~\cite{sliwa2023qc}. 
Even though quantum error correction (QEC) codes are not yet available, quantum supremacy has recently been demonstrated in Google's Sycamore~\cite{arute2019qs}.

The performance of NISQ devices heavily relies on the input problem size, specifically the number of qubits in and the depth of the quantum circuit. 
Noise factors, such as thermal relaxation, can hinder quantum computers (QCs) from executing a high quantum circuit depth \cite{10.1145/3464420}, whereas crosstalk errors reduce parallel gate execution in QCs \cite{acharya2021tp, saki2021as, qian2023am, wilson2020ji, sun2021mr}. 
Researchers have attempted to mitigate these errors using both hardware and software solutions. 
At the software level, a compiler can apply problem-specific optimizations \cite{ding2020sc, murali2020sm, khadirsharbiyani2023tc, xie2021mc, debroy2020lp} tailored to the target hardware. 
At the hardware level, to address crosstalk errors, tunable connections between qubits, achieved using couplers, can be employed.

When using tunable couplers, the connection between qubits is activated only when a two-qubit gate needs execution; otherwise, it remains inactive.
This way, qubits can perform their respective operations without crosstalk errors. 
However, another type of error is introduced. 
Ideally, a SQC is assumed to operate as a closed system at absolute zero temperature, an unattainable condition in the real world. 
Practically, we can only cool the system down to a certain low temperature close to zero, and external interactions are required to control qubits, which introduce errors \cite{lau2022nc}. 
Additionally, the tuning of the coupler (on/off) introduces unwanted interactions and errors to the qubits \cite{Wang2023ct}.

The existing methods~\cite{YANG2023106944, li2020te} focus on optimizing the qubit placement layout for application-specific quantum circuits, aiming to optimize connectivity while considering physical constraints, such as frequency-collision effects, to improve fabrication yield rates. 
However, limited attention has been paid to the configuration of tunable-coupler superconducting quantum computers (SQCs). 
Lin et al. \cite{lin2022ds} proposed a method for finding an optimal layout using tunable couplers. However, this method did not consider the physical constraint imposed by frequency-collision effects. 
Our study bridges the gap by exploring coupling configurations for a specific input quantum circuit. 
It leverages the contextual connectivity correlation of the quantum circuit to minimize connections between noncorrelated qubits (i.e., without two-qubit gates) while enhancing fidelity and considering the physical constraints of the SQC.
We believe our approach can shed light on the benefits of initializing a layout for tunable coupler SQCs instead of dynamically tuning them during runtime, which can significantly affect qubit reliability.

To achieve this, we address several key challenges related to the physical constraints of tunable-coupler QCs. 
Next, we identify the contextual connectivity patterns within the input quantum circuit to discover the optimal connections. 
Afterward, we determine the placement of physical qubits and their connections while ensuring that the final layout complies with the physical constraints of the SQC itself.
\begin{figure}
    \centering
    \begin{subfigure}{0.3\columnwidth}
        \centering
        \includegraphics[width=\columnwidth]{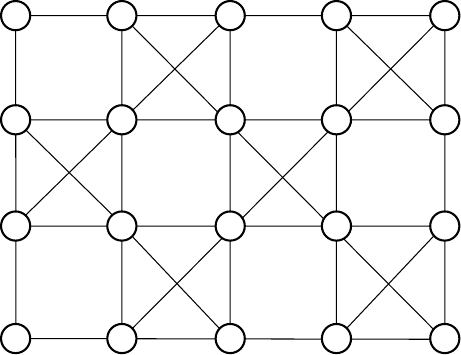}
        \caption{IBM's Q20 Tokyo}
        \label{fig:ibm_tokyo}
    \end{subfigure}
    \hspace{20pt}
    \begin{subfigure}{0.4\columnwidth}
        \centering
        \includegraphics[width=\columnwidth]{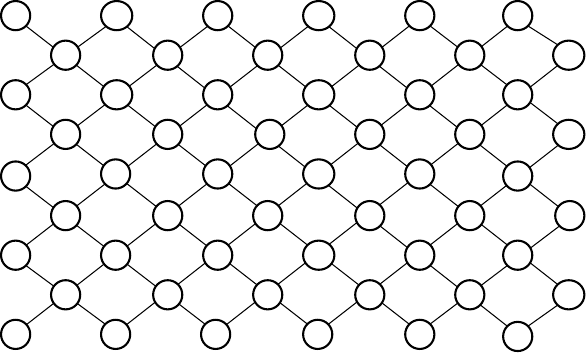}
        \caption{Google's Sycamore}
        \label{fig:google_sycamore}
    \end{subfigure}
    \hfill
    \caption{The IBM Q20 Tokyo and Google Sycamore NISQ devices are represented where each vertex symbolizes a qubit, and each edge signifies a coupler. Therefore, a Controlled-NOT (CNOT) gate can directly execute between any two connected qubits.}
    \label{fig:device_architectures}
\end{figure}
As a solution, we propose a Context-Aware COupler REconfiguration (CA-CORE) method, a compilation flow that considers the connectivity of the input circuit to find the optimal layout configuration. 
We design a circuit analysis that utilizes the correlation matrix between qubits to determine their connectivity weight. 
With this information, we construct a maximum weighted path graph (MWPG) that connects highly correlated qubits. 
Thereafter, we introduce an algorithm to generate a grid graph using the MWPG. Additionally, we propose a heuristic rule for eliminating connections that violate the SQC's constraints on the grid graph. 
In summary, this study makes the following contributions:
\begin{itemize}
    \item We propose CA-CORE, a compilation method for determining the optimal physical qubit-coupling map-layout configuration for a quantum circuit based on tunable coupler-based SQC.
    \item We introduce an algorithm for identifying a MWPG and apply it to construct a grid graph that incorporates both adjacent and diagonal edges while considering the contextual correlations of qubits.
    \item We evaluate the performance of our physical qubit-coupling map-layout configuration, which reduces the depth, number of gates, and SWAP required by up to 18.84\%, 9.48\%, and 42.47\%, respectively. It also improves circuit fidelity by up to 40\% and compiles on a 33-qubit quantum circuit in less than 1$s$.
\end{itemize}

This paper is organized as follows. Section~\ref{sec:background} discusses the background and motivation of our proposed method. The details of the context-aware coupling configuration are discussed in Section~\ref{sec:CA-CORE}. In Section~\ref{sec:evaluation}, we discuss the experimental setup and evaluation results. Section~\ref{sec:discussion} focuses on the discussions and future works, whereas Section~\ref{sec:related_work} presents related works. Finally, the paper is concluded in Section~\ref{sec:conclusion}. 

\section{Background and Motivation}
\label{sec:background}

Here, we provide the fundamentals for understanding our proposed method. We discuss the basics of quantum computing: qubits, quantum operations, and quantum circuits. Thereafter, we discuss the coupler-based SQC hardware constraints, their physical properties, circuit fidelity, and the motivation behind our proposed method.

\begin{figure*}
    \begin{subfigure}[t]{0.3\columnwidth}
        \centering
        \includegraphics[width=\columnwidth]{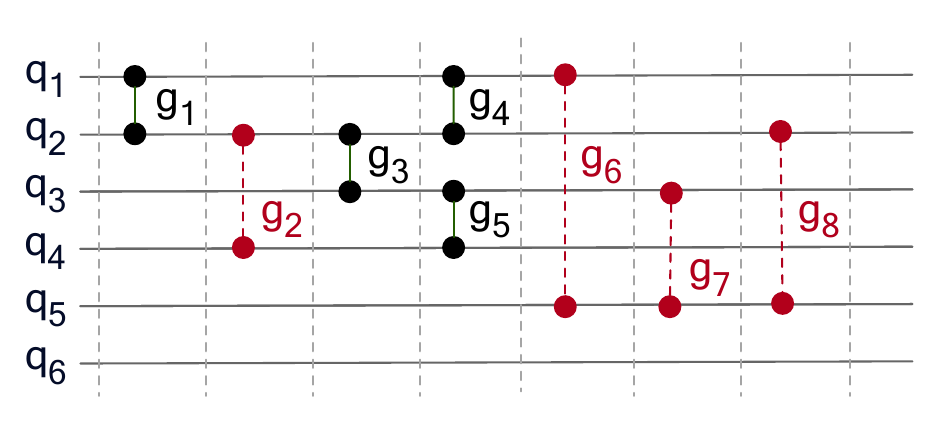}
        \caption{Quantum Circuit}
        \label{fig:original_circuit}
    \end{subfigure}
    \begin{subfigure}[t]{0.4\columnwidth}
        \centering
        \includegraphics[width=\columnwidth]{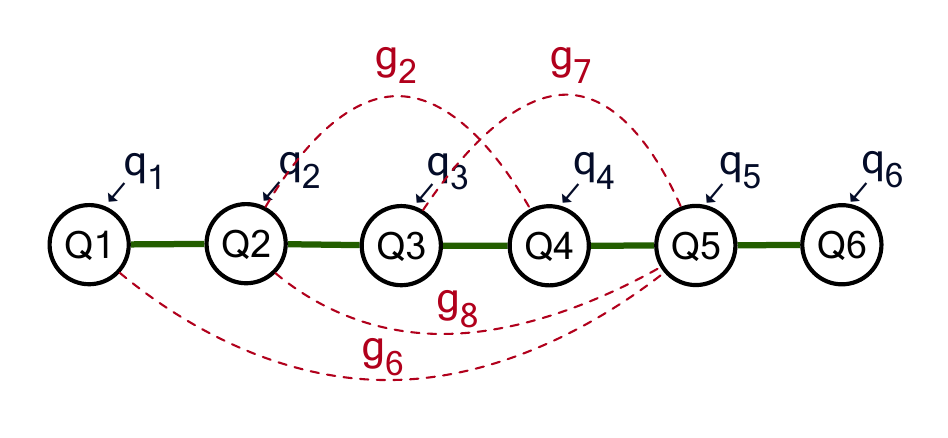}
        \caption{Qubit Mapping on Line Graph}
        \label{fig:lines_graph_circuit}
    \end{subfigure}
    \begin{subfigure}[t]{0.25\columnwidth}
        \centering
        \includegraphics[width=\columnwidth]{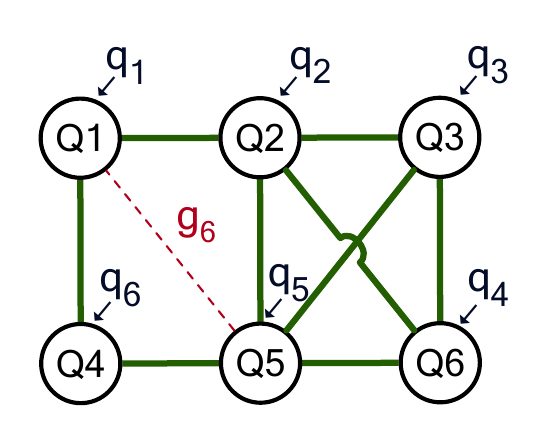}
        \caption{Qubit Mapping on Grid Graph}
        \label{fig:grid_graph_circuit}
    \end{subfigure}
    \caption{Example of a quantum circuit with its corresponding interaction graph (IG). Each line and vertex represents a qubit. The green solid edge indicates that a CNOT gate can be directly executed between the two qubits, and the red dotted edge indicates that a CNOT gate cannot be directly executed between the two qubits.}
    \label{fig:example_circuit}
\end{figure*}
\subsection{Quantum Computing Basics}\label{sec:qubits}
We begin with the basic concepts of quantum bits (qubits) and quantum operations, which were employed to create the quantum program and can be represented using quantum circuits.

\begin{figure}
    \centering
    \begin{subfigure}[t]{0.2\columnwidth}
        \centering
        \includegraphics[width=\columnwidth]{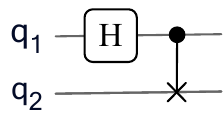}
        \caption{Hadamard and CNOT gate}
        \label{fig:h_cnot_gate}
    \end{subfigure}
    \hspace{20pt}
    \begin{subfigure}[t]{0.4\columnwidth}
        \centering
        \includegraphics[width=\columnwidth]{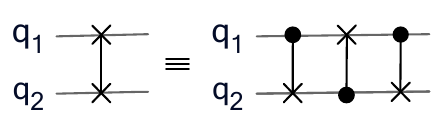}
        \caption{SWAP gate with its equivalent \\ decomposed gates}
        \label{fig:SWAP_gate}
    \end{subfigure}
    \caption{Basic quantum operations. Each horizontal line represents a qubit, and operations (gates) are applied to them, such as Hadamard (H), controlled-NOT (CNOT), SWAP (equivalent to three CNOT) gates.}
    \label{fig:quantum_operations}
\end{figure}

\textbf{Qubits and Quantum Operations}: Quantum programs are composed of qubits and gates as the communication operations~\cite{djordjevic2022qc}. 
Figure~\ref{fig:quantum_operations} shows the quantum basic gates of single and multiple-qubit gates, such as Hadamard (H), Controlled-NOT (CNOT), and SWAP (equivalent to three CNOT) gates. 
A quantum circuit can be composed of as many qubits as needed, denoted as $q$, and represented on horizontal lines in Figure~\ref{fig:original_circuit}. 
The choice of the gates can change the state of the quantum program, and there are two basis states denoted as $|0\rangle$ and $|1\rangle$. 
%
Additionally, SWAP gates can be as expensive as three equivalent CNOT gates. 
We assume that the SWAP gate is considered to have only one weight as the CNOT gate.

\textbf{Quantum Circuit}: An interaction graph (IG) is a graph that represents the interactions between the qubits in a quantum circuit~\cite{siraichi2018qa, li2020te, bandic2023ig}. 
The nodes in the graph represent the qubits $q$, and the edges in the graph represent the two-qubit gates of $q_i$ and $q_j$ as $g_n$, where $n$ is the number of interactions.
To consider a logical circuit, we illustrate the quantum circuit of six qubits ($q_1$ to $q_6$) with the eight CNOT gates ($g_1$ to $g_8$) (see Figure~\ref{fig:original_circuit}). 
For instance, gate $g_1$ represents the direct interaction between qubits $q_1$ and $q_2$, whereas gate $g_2$ represents the indirect interaction between qubits $q_2$ and $q_4$. 
IGs are employed to identify the qubits that can be placed far apart without affecting the circuit's performance of the circuit. Consequently, it helps determine the number of qubit connections in the circuit from the logical to the physical level. This information defines the connectivity weight in this study. 
\subsection{Coupler-Based SQC Hardware Constraints}\label{sec:architectural_constraint}
Quantum computing hardware can be developed using various technologies, such as SQCs, ion traps, quantum dots, neutral atoms, etc. The SQC is currently one of the most promising technology. Shown in Figure~\ref{fig:device_architectures} are the architecture of the IBM's Q20 Tokyo, and Google's Sycamore. During the NISQ era, all logical qubits must be directly implemented by physical qubits unaided by any QEC codes. This study considers two major limitations and constraints of NISQ hardware as context-aware to the model we employed to design coupler reconfiguration for tunable coupler-based SQCs.

\textbf{SQC Physical Properties}: Using two-qubit gates relies on physical connections, meaning superconducting qubits are confined to neighboring locations on a 2D plane.
A popular coupling structure is the layout of qubits on IBM Q20 Tokyo’s superconducting quantum chip, as shown in Figure~\ref{fig:ibm_tokyo}. 
In this coupler structure, there is a major constraint involving the diagonal coupler: continuous diagonal couplers introduce defects of dense frequency collisions on the device. 
Additionally, the vendor fabricates the qubits coupler in a regularized structure of $4 \times 5$ lattices to reduce the fabrication complexity while ensuring the scalability and reliability of the coupling structure.

\textbf{Quantum Circuit Fidelity}: Notably, the previous problem is resolved to optimize qubit connection and maximize the overall performance against topologies by IBM and Google. Li et al.~\cite{li2019tt} reported the average error rates of a single-qubit gate and CNOT gate as $4.43 \times 10^{-3}$ and $3.00 \times 10^{-2}$, respectively. Due to on-chip placement constraints and routing constraints, couplers can only connect one qubit to its neighboring qubit. ~\cite{siraichi2018qa, li2020qm, datta2023ic}. 
Besides the physical properties, we are concerned with the crosstalk error, which affects the quality of circuit fidelity. 
In this study, we investigate the contextual correlation between qubits, i.e., multi-qubit gates. 
We do not consider single-qubit gates because they do not incur additional qubit movement overhead during execution on any coupling structure.
\subsection{Motivation}\label{sec:motivation}
We employ a circuit of six qubits and eight gates to explain our motivation for considering qubit mapping, gate error, and different types of qubit architecture for determining the number of physical qubit coupler reconfiguration for tunable coupler-based SQCs.

\textbf{Quantum circuit's effect on qubit mapping}: When executing a quantum circuit on a QC, the logical qubits need to be mapped to physical qubits on the target quantum architecture.

\textbf{Qubit mapping's impact on the coupling map}: Consider the example of Figure~\ref{fig:lines_graph_circuit}, where the original circuit is mapped onto the line graph with its physical constraints. Besides the indirect interaction of $g_2$, $g_6$, $g_7$ and $g_8$ denotes a red dotted line, other interactions can be mapped directly to the graph. In the same way, Figure~\ref{fig:grid_graph_circuit} maps the same original circuit onto another graph of a grid graph with more physical connectivity. We can clearly observe that the red dotted lines that denote the indirect interactions reduce to only one which is $g_6$, compared with the features of the line graph qubit mapping. 

In this case, the choice of architecture significantly impacts the initial mapping. Hence, bridging the gap between the physical qubit graph and the qubit mapping problem is crucial at the compilation level. We propose a compilation method for determining the maximum physical qubit coupler reconfiguration in tunable coupler-based superconducting quantum computers (SQCs). This method addresses indirect interactions of nonexecutable two-qubit gates and compensates for connectivity limitations. It offers advantages such as reduced circuit depth, fewer SWAP gates, and improved fidelity.
\section{Proposed Context-Aware Coupler Reconfiguration}\label{sec:CA-CORE}

This section covers CA-CORE, our method for context-aware coupling reconfiguration, explaining its architecture and implementation details.

\subsection{Overall Architecture}
\begin{figure*}[!t] 
    \begin{subfigure}[c]{\textwidth}
        \centering
        \includegraphics[width=\textwidth]{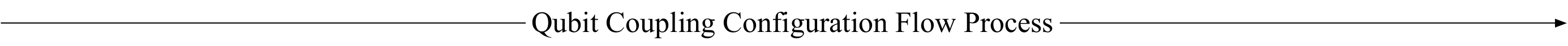}
    \end{subfigure}
    \begin{subfigure}[t]{0.21\textwidth}
        \centering
        \includegraphics[width=\textwidth]{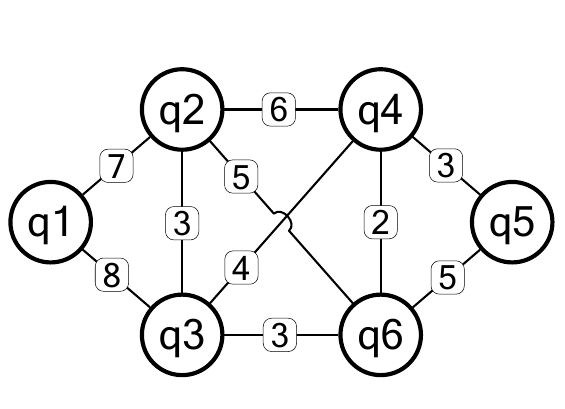}
        \caption{}
        \label{fig:flow_process}
    \end{subfigure}
    \begin{subfigure}[t]{0.21\textwidth}
        \centering
        \includegraphics[width=\textwidth]{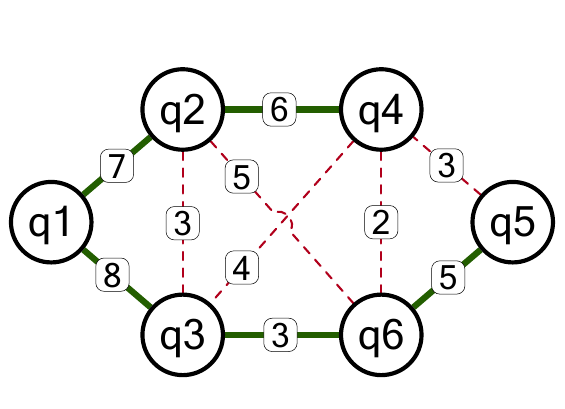}
        \caption{}
        \label{fig:flow_weighted_graph}
    \end{subfigure}
    \begin{subfigure}[t]{0.18\textwidth}
        \centering
        \includegraphics[width=\textwidth]{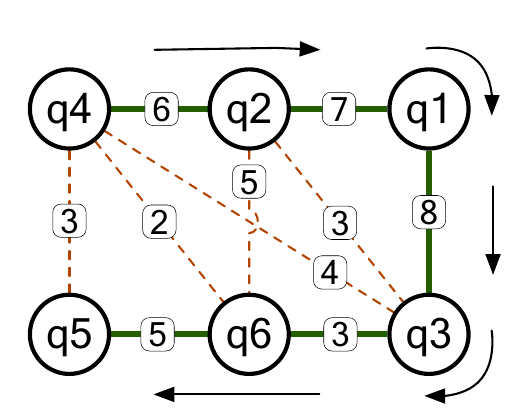}
        \caption{}
        \label{fig:flow_grid_graph}
    \end{subfigure}
    \begin{subfigure}[t]{0.18\textwidth}
        \centering
        \includegraphics[width=\textwidth]{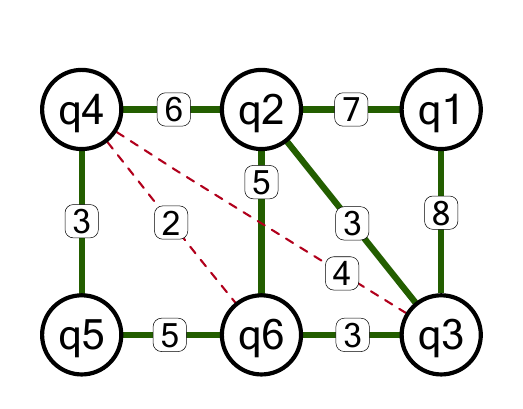}
        \caption{}
        \label{fig:flow_conf_graph}
    \end{subfigure}
    \begin{subfigure}[t]{0.18\textwidth}
        \centering
        \includegraphics[width=\textwidth]{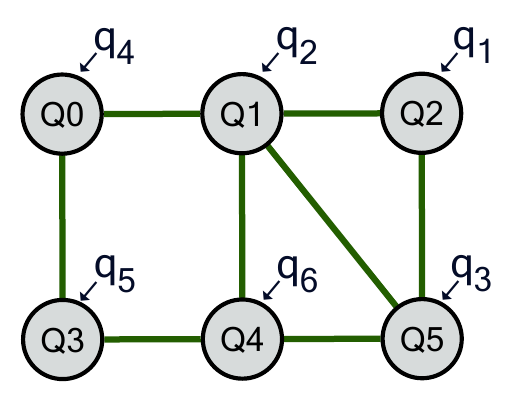}
        \caption{}
        \label{fig:flow_efficient_coupling}
    \end{subfigure}
    \hspace*{\fill}
    \caption{Overview of the coupling map generation process. (a) Analyze the contextual correlation between qubits; the coupling weight is assigned depending on the number of correlations between qubits. (b) Generating the Max Weighted Path Graph (MWPG) by connecting the highest weighted path (solid line). (c) Generating the grid topology using MWPG and connecting adjacent edges. (d) Generating the grid topology using MWPG and connecting diagonal edges. (e) Generated qubit physical layout coupling map.}
    \label{fig:overall_process}
\end{figure*}
\begin{figure}
    \centering
    \includegraphics[width=1\columnwidth]{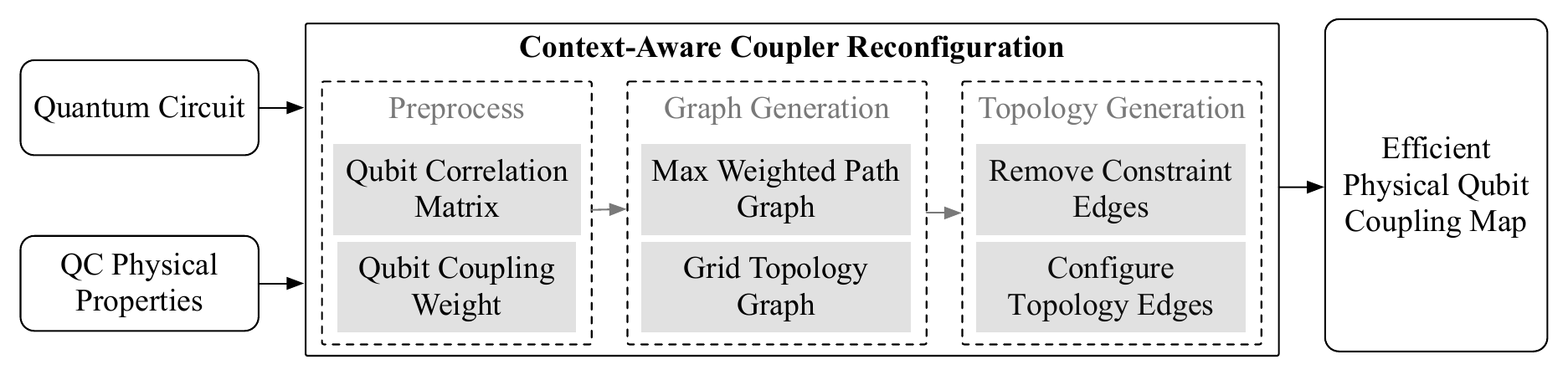}
    \caption{Overall architecture of the proposed method. We load the quantum circuit and QC physical properties as an input to design an efficient physical qubit coupling map.}
    \label{fig:overall_architecture}
\end{figure}
To determine the optimal physical qubit coupling map-layout configuration for a quantum circuit, we must analyze the circuit to establish contextual correlations between logical qubits efficiently. 
This process involves assessing the two-qubit gates between logical qubits, with one-qubit gates being disregarded since they do not require any qubit movement during the execution process. 
For each two-qubit gate, we assign a weight to each qubit pair's connection to create a correlation matrix. 
As illustrated in Figure~\ref{fig:flow_process}, the circuit analysis provides the contextual correlation weights between qubits, with each qubit pair assigned a weight corresponding to the number of two-qubit gates interaction involved in the corresponding qubit pair.

Given that the SQC processor adheres to a 2D grid shape~\cite{arute2019qs, mukai2020p2, crawford2023ca, zhang2023ae, gong2021qw}, we need to reshape the graph from Figure~\ref{fig:flow_process} to conform to a grid-like structure. 
This adjustment should also account for the constraints imposed by the SQC's physical construction, including the transmon qubits and couplers. 
Each qubit can connect to its neighbors in all directions (grid); however, diagonal edges are prohibited when neighboring connections already include diagonal edges (except diagonal neighbor)~\cite{li2020te, nishio2020es}. 

To achieve this, we tailor the circuit graph to minimize qubit movement and distance while maximizing the weight using an MWPG. 
Figure~\ref{fig:flow_weighted_graph} shows an MWPG subject to constraints where each node can only have two edges and loops are disallowed. 
The black (straight) lines represent the maximum weighted path, while the red (dotted) lines indicate edges eliminated for violating the path graph constraints.

We employ this method to construct a path graph while maximizing the edges with the highest weights. 
Subsequently, we transform the MWPG into a grid graph while considering the constraints discussed in Section~\ref{sec:architectural_constraint}.
As shown in Figure~\ref{fig:flow_grid_graph}, this grid graph retains the maximum weighted path and facilitates the connection of adjacent and diagonal edges between nodes. 
Figure~\ref{fig:flow_conf_graph} shows an example where the edges between $q_2$ to $q_6$ and $q_2$ to $q_3$ can be connected, as these connections adhere to the allowed constraints. 
However, the edge between $q_4$ and $q_6$ cannot be connected, as the connection violates the constraints. 
The decision to connect $q_2$ and $q_3$ over $q_4$ and $q_6$ is based on the edge weights, with the former having a weight of 3 and the latter having a weight of 2.

After completing these processes, the grid graph constructed in Figure~\ref{fig:flow_conf_graph} can be employed to establish a physical qubit-coupling map, as shown in Figure~\ref{fig:flow_efficient_coupling}. 
All allowed edges are utilized to form the couplers for the qubits. 
The result is an efficient qubit coupling map for the input circuit. 
Thus far, we have presented an overview of the entire process and the objectives of our proposed method using a small quantum circuit with only six qubits. To delve into the details of each implementation step, particularly for large quantum circuits, we can refer to the overall proposed architecture described in Figure~\ref{fig:overall_architecture}.

The algorithm begins the prepossess by analyzing the input quantum circuit to extract the coupling map (edges) and its correlation weight matrix.
The coupling map is established by detecting connections between qubits via two-qubit gates. 
Whenever two qubits are connected in this manner, they are considered a pair or a coupling. 
Meanwhile, for the correlation weight matrix, we assign a weight of 1 to each two-qubit gate present in the quantum circuit.

Using contextual results obtained from the analysis of the coupling map and the correlation matrix, we can subsequently proceed to configure an efficient qubit coupling map layout.
\begin{algorithm}[!t]
    \KwIn{\\
        $qc\_graph$: original quantum circuit graph, \\
        $c\_weight$: qubit coupling weight correlation
    }
    \KwOut{\\
        $path\_graph$ : max weighted path graph  
    }
    \SetKwProg{generate}{Function \emph{}}{}{end}

    \generate{generate\_max\_weighted\_path\_graph}{
        $path\_graph$ = \text{initial empty graph};
        

        $sorted\_edge$ = get sort weighted edges in $c\_weight$;

                
        \For{$edge$ \textbf{in} $sorted\_edge$}{
            \tcp{extract info from edge}
            $n1$, $n2$, $weight$ = $edge$; 
            
            \tcp{max connection validation}
            \If{$n1.neighbors >= 2$ or $n2.neighbors >= 2$}{
                \tcp{only allow max 2 edges}
                continue; 
            }

            \tcp{loop connection validation}
            $tmp\_path\_graph$ = $path\_graph$;\\
            \textbf{add} $edge(n1,n2)$ \textbf{to} $tmp\_path\_graph$;\\
            \If{$tmp\_path\_graph.has\_loop\_subgraph$}{
                \tcp{loop is not allowed}
                continue; 
            }

            \textbf{add} $edge(n1,n2,weight)$ \textbf{to} $path\_graph$;\\

        }
    \Return $path\_graph$;
    }
    
    \caption{Max Weighted Path Graph Generation}
    \label{alg:max_weighted_path_gen}
\end{algorithm}
\subsection{Max Weighted Path Graph Generation}
\label{sec:max_weighted_path} 

As we discussed in the previous section, the topology of the quantum circuit resembles a grid, featuring both adjacent and diagonal edges. 
Consequently, the configured qubit coupling map topology should follow this grid-like format. 
To achieve an efficient grid-like layout, we initiate the process by generating an MWPG using the logical coupling map and correlation data obtained from the analyzed circuit. 
The MWPG is employed because it can be efficiently transformed into the desired grid-like structure with a specified number of rows and columns. 
Moreover, it can accommodate qubit layouts characterized by high correlations between individual qubits.

Algorithm~\ref{alg:max_weighted_path_gen} outlines the procedure for generating an MWPG based on the logical qubit-coupling weight correlations defined as \textit{c\_weight}.
There are two conditions that an MWPG must adhere to, as follows: 
1) A single logical qubit \textit{n} cannot have more than 2 edges or \textit{neighbors} connections with other qubits. 
2) Loops are not permitted; to effect this, we make a temporary path graph (\textit{tmp\_path\_graph}) with all the edges and nodes from the original \textit{path\_graph}. Thereafter, we add the edges with two qubits (\textit{n1} and \textit{n2}) to \textit{tmp\_path\_graph}. 
A subsequent verification process is then carried out to detect the presence of any loops within this temporary graph.
In this process, logical couplings with the highest weights are constructed first, followed by those with lower weights. 
If a coupling fails to meet the mentioned conditions, no coupling connection will form. 

An issue may emerge based on the contextual correlation of the quantum circuit when certain qubits lack of connection with others. 
In such instances, our MWPG might contain multiple isolated subgraphs. 
To resolve this, we connect these subgraphs by identifying logical qubits with a single connection, maintaining the prior constraints. 



\begin{algorithm} []
    \KwIn{\\
        $nrow$: number of rows for grid topology,\\
        $ncol$: number of columns for grid topology,\\
        $c\_weight$: qubit coupling weight correlation,\\
        $path\_graph$: max weighted path graph\\
    }
    \KwOut{\\
        $grid\_graph$ : grid graph with adjacency and diagonal edges
    }
    \SetKwProg{generate}{Function \emph{}}{}{end}

    \generate{generate\_grid\_topology\_graph}{
        \tcp{assign node to $pos\_matrix$ to determine the position of the node in the grid graph}
        $pos\_matrix$ = put $path\_graph$'s node by shape $row$ x $col$;\\
        
        $grid\_graph$ = get all nodes and edges from $path\_graph$;\\

        \tcp{connect adjacent edges}
        \For{$row \gets 0$ \textbf{to} $nrow-1$}{
            \For{$col \gets 0$ \textbf{to} $ncol$}{
                $n1$ = $pos\_matrix[row][col]$;\\
                $n2$ = $pos\_matrix[row+1][col]$;\\

                \If{$not$ $c\_weight.has\_edge(n1, n2)$}{
                    continue; 
                }
                $grid\_graph.add(n1,n2)$
            }
        }
        \tcp{connect diagonal edges}
        \For{$row \gets 0$ \textbf{to} $nrow-1$}{
            \For{$col \gets 0$ \textbf{to} $ncol$}{
                $n1$ = $pos\_matrix[row][col]$;\\
                \tcp{left lower diagonal edge}
                \If{$col > 0$}{
                    $n2\_left$ = $pos\_matrix[row+1][col-1]$;\\
                    \If{$not$ $c\_weight.has\_edge(n1,n2\_left)$}{
                        continue;
                    }
                    $grid\_graph.add(n1, n2\_left$
                }
                \tcp{right lower diagonal edge}
                \If{$col+1 < ncol$}{
                    $n2\_right$ = $pos\_matrix[row+1][col+1]$;\\
                    \If{$not$ $c\_weight.has\_edge(n1,n2\_right)$}{
                        continue;
                    }
                    
                    $grid\_graph.add(n1,n2\_right)$
                }
            }
        }

    \Return $grid\_graph$;
    }
    \caption{Grid Topology Generation}
    \label{alg:grid_topology_generation}
\end{algorithm}
\begin{figure}[!t]
    \centering{\includegraphics[width=6cm]{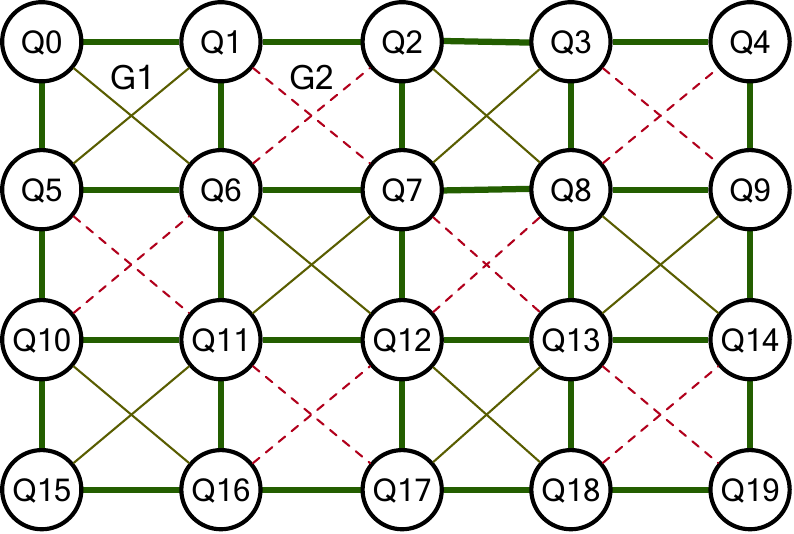}}
    \caption{Physical qubit grid topology graph configuration with all possible edges. Diagonal edges are separated into G1: solid (black) line and G2: dotted (red) line. Each group represents possible diagonal edge configurations that conform to the maximum diagonal edges allowed while considering to the physical constraints of SQC.}
    \label{fig:constraint_configure}
\end{figure}
\subsection{Grid Topology Generation}

In Figures~\ref{fig:flow_weighted_graph} and~\ref{fig:flow_grid_graph}, the grid graph is created by arranging nodes adjacent to each other in rows and columns.
To initialize the grid graph to construct our proposed MWPG, we select either the first or last node of the path and position them in a row and column format. Thereafter, we connect it to all of the path nodes through edges. 

To implement this in a programming model, as shown in Algorithm~\ref{alg:grid_topology_generation}, we utilize a position matrix (\textit{pos\_matrix}) to determine the position of each logical qubit by placing its node position from MWPG in the matrix. 
Afterward, we connect the logical qubits in adjacent positions only if their qubit coupling-weight (\textit{c\_weight}) are correlated. 
The adjacent edge can be formed using \textit{pos\_matrix} by iterating through each row and column (\textit{col}). 
In each col, the position of \textit{n1} in the \textit{pos\_matrix} is located at index \textit{pos\_matrix[row][col]}. 
Meanwhile, its adjacent edge, \textit{n2}, is positioned at index \textit{pos\_matrix} \textit{[row+1][col]}. 
Essentially, the adjacent edge of \textit{n1} is situated at its bottom in the grid graph. 
As depicted in Figure~\ref{fig:flow_grid_graph}, \textit{n1} is represented by the \textit{q4} in the first row and first column, whereas its adjacent edge, \textit{n2}, is represented by \textit{q5} in the second row and first column.

For the diagonal edges, we initiate the connection process starting from the first row and first column. 
Each logical qubit in the first row checks the availability of its left and right diagonal edges through the \textit{pos\_matrix}, constructing an edge only when the \textit{c\_weight} values are correlated.
The left diagonal edge of \textit{n1}, referred to as \textit{n2\_left}, can be located at the bottom-left of its position within the \textit{grid\_graph}, with index \textit{pos\_matrix[row+1][col-1]}. 
It exists only when the condition, $col > 0$ is satisfied, meaning that \textit{n2\_left} is absent if \textit{n1} is positioned at the extreme left of the \textit{grid\_graph}. 
As for the right diagonal edge, denoted as \textit{n2\_right}, it can be found at index \textit{pos\_matrix[row+1][col+1]}. 
It exists only when the condition $col + 1 < ncol$ is met, signifying that there is no \textit{n2\_right} if \textit{n1} is at the far right of the \textit{grid\_graph}. 
For an example, please refer to Figure~\ref{fig:flow_conf_graph}, where \textit{q2, q5, and q3} represent \textit{n1, n2\_left, n2\_right}, respectively.

Upon connecting all correlated adjacency and diagonal edges, we obtain a grid graph that maximizes connections for logical qubits, ensuring that all possible edges are established. 
This minimizes travel distances between correlated qubits, while minimizing unnecessary coupling connections. 
Notably, throughout this process, we did not considered the physical constraints associated with diagonal edge connections. 
This concern will be addressed more precisely in the following section.

\subsection{Eliminating Constraints on Diagonal Edges} \label{sec:topology_diagonal_edges_configuration} 

To address the constraints imposed by the physical qubits of the SQC, as discussed in Section~\ref{sec:architectural_constraint}, we must consider which possible configuration of diagonal edges carries greater weight. 
To achieve this, we partition the diagonal edges into two groups, each accumulating the combined weight of diagonal edges that can coexist within the grid graph. 
As shown in Figure~\ref{fig:constraint_configure}, a grid graph with all possible edges connected is depicted. 
The diagonal edges are separated into two groups: Group 1 (G1) and Group 2 (G2). G1 contains all the diagonal edges with solid (black) lines, whereas G2 comprises all the diagonal edges with dotted (red) lines. 
In CA-CORE, such a reconfigured topology graph is only generated when the quantum circuit contains fully connected qubits, meaning each qubit correlates with all the other qubits in the quantum circuit.

We begin accumulating the weight of the edges for G1, starting from the first row and first column. 
Each group within each row and column consists of two pairs of diagonal edges between physical qubits, with the weight of each edge being added to the group. 
Subsequently, moving to the first row and the second column of physical qubits, another two pairs of diagonal edges to the right can be accumulated. 
However, this time, the accumulated weight belongs to G2. 
This process continues until no more neighboring qubits are to the right of a given position.
Subsequently, the group orientation needs to be flipped in the second row, as demonstrated in Figure \ref{fig:constraint_configure}.

Finally, we eliminate the diagonal edges from the group that has accumulated the lowest weight. 
The resulting grid graph serves as the basis for constructing the layout of the physical qubit topology. 
This topology layout maximizes the correlation weight of the logical qubits while adhering to the constraints of the SQC itself.
\section{Evaluation} \label{sec:evaluation}
In this section, we present the evaluation of CA-CORE compared with various existing topologies, in terms of depth, gate, and SWAP-gate-number reduction, and fidelity improvement.
\subsection{Experiment Setup}
\label{sec:experiment_setup}
\textbf{Benchmarks.} We generate random quantum circuits using Qiskit \cite{Qiskit}, ranging from 10-qubit to 33-qubit configurations. These circuits span up to approximately 200 depths and 3000 gates. This allows us to assess the performance regarding depth reductions, number of gates, and SWAP operations. Furthermore, we utilize the low-level QASM benchmarks from~\cite{li2022qasmbench}, which are well-suited for NISQ evaluation and simulation. These benchmarks aid in assessing noise-tolerant performance, given their considerable depth. Moreover, their expectation values can be simulated classically, enabling the extraction of precise results.

\textbf{Hardware Topology.} We compare the compiled CA-CORE with the existing diversity of fixed-coupling device topologies, such as `ibmq\_almaden,' `ibmq\_cairo,' and `ibmq\_prague' available on Qiskit~\cite{Qiskit}. Furthermore, we generated a topology inspired by Google's Sycamore architecture~\cite{arute2019qs}.

\textbf{Experiment Platform.} All experiments presented in this paper were conducted on a server with an AMD Ryzen 5955WX (32 logical cores) CPU, 512GB of memory, and 3 Nvidia RTX 4090 GPUs.

\textbf{Algorithm Configuration.} Our proposed method was set up to compile circuits to a topology where the number of physical qubits matches the number of logical qubits in the quantum circuit. Couplers are configured only when a correlation (a two-qubit gate) between logical qubits exists. The grid size is adjusted based on the required number of qubits.

\subsubsection{Depth, Gate, and SWAP-gate-number Reduction} \label{sec:eval_SWAP}

To assess the performance of our CA-CORE in terms of depth, gate, and SWAP-gate-number reduction, we conducted experiments using 10 different randomly generated quantum circuits for each qubit count, as mentioned in Section~\ref{sec:experiment_setup} of our setup benchmarks. 
This approach aims to evaluate the diversity of the contextual qubit correlation, as opposed to using practical circuits that typically exhibit specific patterns in their qubit correlations.

To determine the number of gates, depth, and SWAP operations for all topologies mentioned in Section~\ref{sec:experiment_setup}, including CA-CORE, we utilize Qiskit's transpiler. 
We configure a custom coupling map setup and set the \textit{optimization\_level} to $0$. 
In this configuration, Qiskit applies minimal optimization, primarily focusing on trivial qubit mapping, where logical qubit 0 is mapped to physical qubit 0, and so forth. 
Subsequently, SWAP gates are inserted to establish connectivity for logical qubits that are not directly connected.
\begin{table*}[t]
\scalebox{0.57}{
\begin{tabular}{|c|ccc|ccc|ccc|ccc|}
\hline
\multirow{2}{*}{\textbf{\begin{tabular}[c]{@{}c@{}}Num \\ of Qubit\end{tabular}}} & \multicolumn{3}{c|}{\textbf{CA-CORE to Almaden (20)}}                                               & \multicolumn{3}{c|}{\textbf{CA-CORE to Cairo (27)}}& \multicolumn{3}{c|}{\textbf{CA-CORE to Prague (33)}}                                                & \multicolumn{3}{c|}{\textbf{CA-CORE to Sycamore (53)}}
\\ \cline{2-13} 
                                                                                 & \multicolumn{1}{c|}{\textbf{Depth}}   & \multicolumn{1}{c|}{\textbf{Gate}}   & \textbf{SWAP}    & \multicolumn{1}{c|}{\textbf{Depth}}   & \multicolumn{1}{c|}{\textbf{Gate}}   & \textbf{SWAP}    & \multicolumn{1}{c|}{\textbf{Depth}}   & \multicolumn{1}{c|}{\textbf{Gate}}   & \textbf{SWAP}    & \multicolumn{1}{c|}{\textbf{Depth}}  & \multicolumn{1}{c|}{\textbf{Gate}}   & \textbf{SWAP}    \\ \hline\hline
10                                                                               & \multicolumn{1}{c|}{10.54\%}          & \multicolumn{1}{c|}{4.23\%}          & 35.09\%          & \multicolumn{1}{c|}{10.02\%}          & \multicolumn{1}{c|}{8.04\%}          & 51.69\%          & \multicolumn{1}{c|}{10.68\%}          & \multicolumn{1}{c|}{8.09\%}          & 51.87\%          & \multicolumn{1}{c|}{6.60\%}          & \multicolumn{1}{c|}{5.01\%}          & 39.21\%          \\ \hline
11                                                                               & \multicolumn{1}{c|}{10.66\%}          & \multicolumn{1}{c|}{5.55\%}          & 41.35\%          & \multicolumn{1}{c|}{13.03\%}          & \multicolumn{1}{c|}{9.02\%}          & 54.31\%          & \multicolumn{1}{c|}{13.28\%}          & \multicolumn{1}{c|}{9.22\%}          & 54.92\%          & \multicolumn{1}{c|}{7.92\%}          & \multicolumn{1}{c|}{5.89\%}          & 42.89\%          \\ \hline
12                                                                               & \multicolumn{1}{c|}{11.45\%}          & \multicolumn{1}{c|}{5.57\%}          & 39.04\%          & \multicolumn{1}{c|}{13.39\%}          & \multicolumn{1}{c|}{8.71\%}          & 50.88\%          & \multicolumn{1}{c|}{12.97\%}          & \multicolumn{1}{c|}{8.43\%}          & 49.98\%          & \multicolumn{1}{c|}{8.58\%}          & \multicolumn{1}{c|}{5.56\%}          & 39.00\%          \\ \hline
13                                                                               & \multicolumn{1}{c|}{12.51\%}          & \multicolumn{1}{c|}{5.50\%}          & 35.59\%          & \multicolumn{1}{c|}{12.68\%}          & \multicolumn{1}{c|}{8.26\%}          & 46.10\%          & \multicolumn{1}{c|}{13.76\%}          & \multicolumn{1}{c|}{8.75\%}          & 47.67\%          & \multicolumn{1}{c|}{9.01\%}          & \multicolumn{1}{c|}{6.36\%}          & 39.20\%          \\ \hline
14                                                                               & \multicolumn{1}{c|}{10.42\%}          & \multicolumn{1}{c|}{4.83\%}          & 32.32\%          & \multicolumn{1}{c|}{11.28\%}          & \multicolumn{1}{c|}{7.53\%}          & 43.42\%          & \multicolumn{1}{c|}{12.53\%}          & \multicolumn{1}{c|}{7.89\%}          & 44.66\%          & \multicolumn{1}{c|}{7.54\%}          & \multicolumn{1}{c|}{5.53\%}          & 35.52\%          \\ \hline
15                                                                               & \multicolumn{1}{c|}{11.59\%}          & \multicolumn{1}{c|}{4.53\%}          & 30.17\%          & \multicolumn{1}{c|}{13.51\%}          & \multicolumn{1}{c|}{7.25\%}          & 41.59\%          & \multicolumn{1}{c|}{14.75\%}          & \multicolumn{1}{c|}{7.86\%}          & 43.73\%          & \multicolumn{1}{c|}{10.18\%}         & \multicolumn{1}{c|}{6.02\%}          & 36.86\%          \\ \hline
16                                                                               & \multicolumn{1}{c|}{10.76\%}          & \multicolumn{1}{c|}{4.86\%}          & 30.53\%          & \multicolumn{1}{c|}{15.07\%}          & \multicolumn{1}{c|}{7.85\%}          & 42.29\%          & \multicolumn{1}{c|}{14.22\%}          & \multicolumn{1}{c|}{7.41\%}          & 40.79\%          & \multicolumn{1}{c|}{8.23\%}          & \multicolumn{1}{c|}{4.80\%}          & 30.27\%          \\ \hline
17                                                                               & \multicolumn{1}{c|}{13.11\%}          & \multicolumn{1}{c|}{5.00\%}          & 29.77\%          & \multicolumn{1}{c|}{17.04\%}          & \multicolumn{1}{c|}{7.94\%}          & 41.00\%          & \multicolumn{1}{c|}{16.74\%}          & \multicolumn{1}{c|}{7.59\%}          & 39.82\%          & \multicolumn{1}{c|}{10.46\%}         & \multicolumn{1}{c|}{5.84\%}          & 33.33\%          \\ \hline
18                                                                               & \multicolumn{1}{c|}{11.59\%}          & \multicolumn{1}{c|}{5.52\%}          & 31.74\%          & \multicolumn{1}{c|}{17.42\%}          & \multicolumn{1}{c|}{8.65\%}          & 43.00\%          & \multicolumn{1}{c|}{17.72\%}          & \multicolumn{1}{c|}{9.00\%}          & 44.06\%          & \multicolumn{1}{c|}{9.13\%}          & \multicolumn{1}{c|}{5.74\%}          & 32.65\%          \\ \hline
19                                                                               & \multicolumn{1}{c|}{13.34\%}          & \multicolumn{1}{c|}{5.45\%}          & 29.86\%          & \multicolumn{1}{c|}{19.92\%}          & \multicolumn{1}{c|}{9.22\%}          & 42.87\%          & \multicolumn{1}{c|}{18.60\%}          & \multicolumn{1}{c|}{9.27\%}          & 43.00\%          & \multicolumn{1}{c|}{7.87\%}          & \multicolumn{1}{c|}{5.08\%}          & 28.34\%          \\ \hline
20                                                                               & \multicolumn{1}{c|}{12.18\%}          & \multicolumn{1}{c|}{4.91\%}          & 26.07\%          & \multicolumn{1}{c|}{19.01\%}          & \multicolumn{1}{c|}{9.44\%}          & 41.59\%          & \multicolumn{1}{c|}{18.33\%}          & \multicolumn{1}{c|}{9.90\%}          & 42.86\%          & \multicolumn{1}{c|}{6.67\%}          & \multicolumn{1}{c|}{4.88\%}          & 25.93\%          \\ \hline
21                                                                               & \multicolumn{1}{c|}{-}                & \multicolumn{1}{c|}{-}               & -                & \multicolumn{1}{c|}{19.36\%}          & \multicolumn{1}{c|}{9.24\%}          & 39.99\%          & \multicolumn{1}{c|}{18.82\%}          & \multicolumn{1}{c|}{9.14\%}          & 39.70\%          & \multicolumn{1}{c|}{6.20\%}          & \multicolumn{1}{c|}{3.98\%}          & 21.34\%          \\ \hline
22                                                                               & \multicolumn{1}{c|}{-}                & \multicolumn{1}{c|}{-}               & -                & \multicolumn{1}{c|}{17.94\%}          & \multicolumn{1}{c|}{9.17\%}          & 39.17\%          & \multicolumn{1}{c|}{19.64\%}          & \multicolumn{1}{c|}{9.60\%}          & 40.36\%          & \multicolumn{1}{c|}{5.25\%}          & \multicolumn{1}{c|}{3.70\%}          & 19.67\%          \\ \hline
23                                                                               & \multicolumn{1}{c|}{-}                & \multicolumn{1}{c|}{-}               & -                & \multicolumn{1}{c|}{19.50\%}          & \multicolumn{1}{c|}{9.92\%}          & 39.99\%          & \multicolumn{1}{c|}{18.95\%}          & \multicolumn{1}{c|}{10.32\%}         & 41.04\%          & \multicolumn{1}{c|}{7.63\%}          & \multicolumn{1}{c|}{4.54\%}          & 22.33\%          \\ \hline
24                                                                               & \multicolumn{1}{c|}{-}                & \multicolumn{1}{c|}{-}               & -                & \multicolumn{1}{c|}{20.37\%}          & \multicolumn{1}{c|}{9.67\%}          & 38.77\%          & \multicolumn{1}{c|}{18.46\%}          & \multicolumn{1}{c|}{9.25\%}          & 37.60\%          & \multicolumn{1}{c|}{8.48\%}          & \multicolumn{1}{c|}{3.92\%}          & 19.45\%          \\ \hline
25                                                                               & \multicolumn{1}{c|}{-}                & \multicolumn{1}{c|}{-}               & -                & \multicolumn{1}{c|}{18.58\%}          & \multicolumn{1}{c|}{8.45\%}          & 34.05\%          & \multicolumn{1}{c|}{20.17\%}          & \multicolumn{1}{c|}{8.87\%}          & 35.26\%          & \multicolumn{1}{c|}{7.05\%}          & \multicolumn{1}{c|}{3.54\%}          & 17.05\%          \\ \hline
26                                                                               & \multicolumn{1}{c|}{-}                & \multicolumn{1}{c|}{-}               & -                & \multicolumn{1}{c|}{21.10\%}          & \multicolumn{1}{c|}{9.62\%}          & 38.81\%          & \multicolumn{1}{c|}{20.81\%}          & \multicolumn{1}{c|}{9.70\%}          & 39.03\%          & \multicolumn{1}{c|}{7.44\%}          & \multicolumn{1}{c|}{4.44\%}          & 21.68\%          \\ \hline
27                                                                               & \multicolumn{1}{c|}{-}                & \multicolumn{1}{c|}{-}               & -                & \multicolumn{1}{c|}{20.34\%}          & \multicolumn{1}{c|}{9.01\%}          & 34.87\%          & \multicolumn{1}{c|}{19.27\%}          & \multicolumn{1}{c|}{9.41\%}          & 35.98\%          & \multicolumn{1}{c|}{3.40\%}          & \multicolumn{1}{c|}{2.59\%}          & 12.58\%          \\ \hline
28                                                                               & \multicolumn{1}{c|}{-}                & \multicolumn{1}{c|}{-}               & -                & \multicolumn{1}{c|}{-}                & \multicolumn{1}{c|}{-}               & -                & \multicolumn{1}{c|}{21.50\%}          & \multicolumn{1}{c|}{10.68\%}         & 40.02\%          & \multicolumn{1}{c|}{6.75\%}          & \multicolumn{1}{c|}{3.93\%}          & 18.59\%          \\ \hline
29                                                                               & \multicolumn{1}{c|}{-}                & \multicolumn{1}{c|}{-}               & -                & \multicolumn{1}{c|}{-}                & \multicolumn{1}{c|}{-}               & -                & \multicolumn{1}{c|}{22.22\%}          & \multicolumn{1}{c|}{10.67\%}         & 39.56\%          & \multicolumn{1}{c|}{6.41\%}          & \multicolumn{1}{c|}{3.69\%}          & 17.36\%          \\ \hline
30                                                                               & \multicolumn{1}{c|}{-}                & \multicolumn{1}{c|}{-}               & -                & \multicolumn{1}{c|}{-}                & \multicolumn{1}{c|}{-}               & -                & \multicolumn{1}{c|}{25.23\%}          & \multicolumn{1}{c|}{11.57\%}         & 40.82\%          & \multicolumn{1}{c|}{8.62\%}          & \multicolumn{1}{c|}{4.46\%}          & 19.74\%          \\ \hline
31                                                                               & \multicolumn{1}{c|}{-}                & \multicolumn{1}{c|}{-}               & -                & \multicolumn{1}{c|}{-}                & \multicolumn{1}{c|}{-}               & -                & \multicolumn{1}{c|}{24.50\%}          & \multicolumn{1}{c|}{11.11\%}         & 39.62\%          & \multicolumn{1}{c|}{7.78\%}          & \multicolumn{1}{c|}{3.83\%}          & 17.30\%          \\ \hline
32                                                                               & \multicolumn{1}{c|}{-}                & \multicolumn{1}{c|}{-}               & -                & \multicolumn{1}{c|}{-}                & \multicolumn{1}{c|}{-}               & -                & \multicolumn{1}{c|}{23.40\%}          & \multicolumn{1}{c|}{11.12\%}         & 38.70\%          & \multicolumn{1}{c|}{5.06\%}          & \multicolumn{1}{c|}{3.48\%}          & 15.39\%          \\ \hline
33                                                                               & \multicolumn{1}{c|}{-}                & \multicolumn{1}{c|}{-}               & -                & \multicolumn{1}{c|}{-}                & \multicolumn{1}{c|}{-}               & -                & \multicolumn{1}{c|}{27.07\%}          & \multicolumn{1}{c|}{12.75\%}         & 42.77\%          & \multicolumn{1}{c|}{7.58\%}          & \multicolumn{1}{c|}{3.60\%}          & 16.04\%          \\ \hline\hline
\textbf{Average:}                                                                & \multicolumn{1}{c|}{\textbf{11.65\%}} & \multicolumn{1}{c|}{\textbf{5.09\%}} & \textbf{32.87\%} & \multicolumn{1}{c|}{\textbf{16.64\%}} & \multicolumn{1}{c|}{\textbf{8.72\%}} & \textbf{42.47\%} & \multicolumn{1}{c|}{\textbf{18.48\%}} & \multicolumn{1}{c|}{\textbf{9.48\%}} & \textbf{42.24\%} & \multicolumn{1}{c|}{\textbf{7.49\%}} & \multicolumn{1}{c|}{\textbf{4.60\%}} & \textbf{25.91\%} \\ \hline
\end{tabular}}
\caption{Comparison of the number of depth, gate, and SWAP-gate-number reduction performances between our compiled CA-CORE and various existing topologies, including IBM's ibmq\_almaden 20-qubit, ibmq\_cairo 27-qubit, and ibmq\_prague 33-qubit systems, as well as Google's Sycamore 53-qubit system. Ten randomly generated quantum circuits with approximately 2000-gate and 200-depth on average were evaluated for each qubit count, and their average results were computed individually.}
\label{tb:SWAP_reduction}
\end{table*}
Table~\ref{tb:SWAP_reduction} presents the performance comparison results between CA-CORE and all benchmark topologies regarding depth, gate, and SWAP-gate-number reductions.
On average, CA-CORE demonstrates improvements of $13.56\%$, $6.97\%$, and $35.87\%$ for depth, number of gates, and SWAP operations, respectively. 
Specifically, for the 30-qubit circuit, CA-CORE employs the coupling-map configuration depicted in Figure~\ref{fig:random_circuit_30_topology}. 
In this setup, CA-CORE utilizes diagonal couplings, resulting in reductions of $8.62\%$, $4.46\%$, and $17.30\%$ for depth, number of gates, and SWAP counts, respectively, in contrast to the result of the Sycamore coupling map, as illustrated in Figure~\ref{fig:google_sycamore}. 
The primary distinction between the two configurations lies in the additional diagonal couplings among qubits, demonstrating that integrating these essential coupler connections significantly enhances the performance.

Regarding the gate and depth-reduction count, in our analysis, we treat each SWAP gate as a single gate, and we do not consider it as a three-CNOT gate decomposition. 
This means that a SWAP gate is counted as one gate when summing all the gates in the circuit. 
The depth count reflects the level of efficient parallel execution and reduction in SWAP path overhead within the quantum system, considering its topological setup.
\begin{figure*}[!t]
    \begin{subfigure}[t]{0.32\columnwidth}
        \centering
        \caption{bigadder\_n18' Circuit}
        \includegraphics[width=\columnwidth]{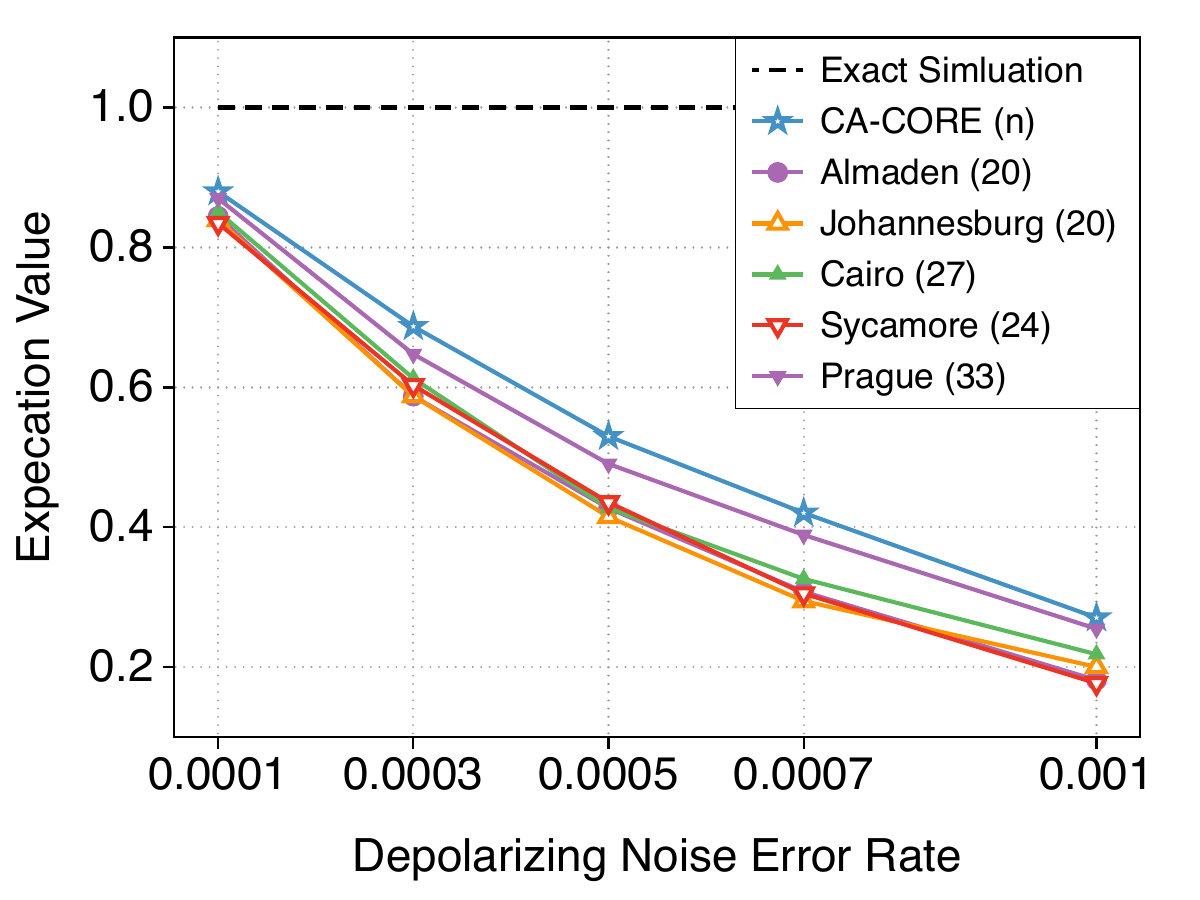}
        \label{fig:bigadder}
    \end{subfigure}
    \begin{subfigure}[t]{0.32\columnwidth}
        \centering
        \caption{qec9xz\_n17' Circuit}
        \includegraphics[width=\columnwidth]{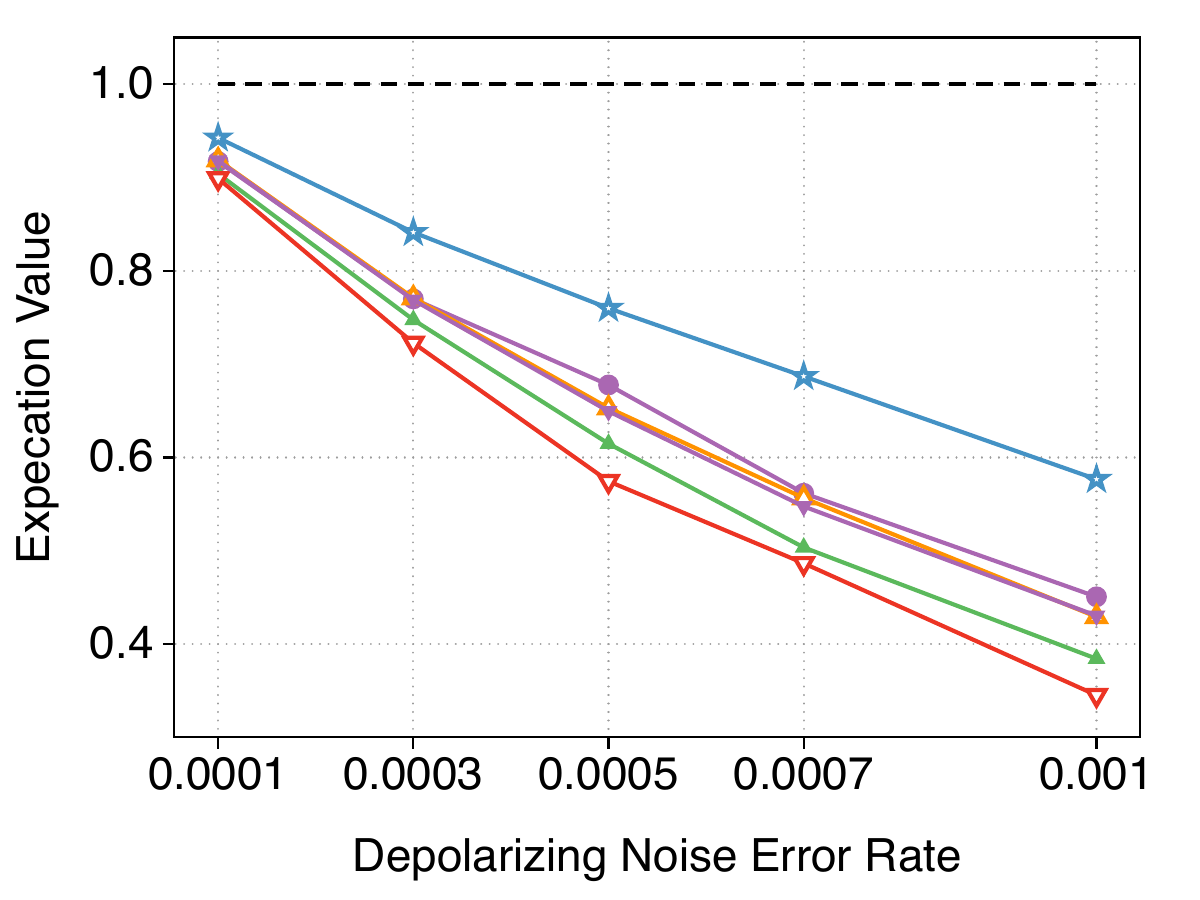}
        \label{fig:qec9xz}
    \end{subfigure}
    \begin{subfigure}[t]{0.32\columnwidth}
        \centering
        \caption{bv\_n14' Circuit}
        \includegraphics[width=\columnwidth]{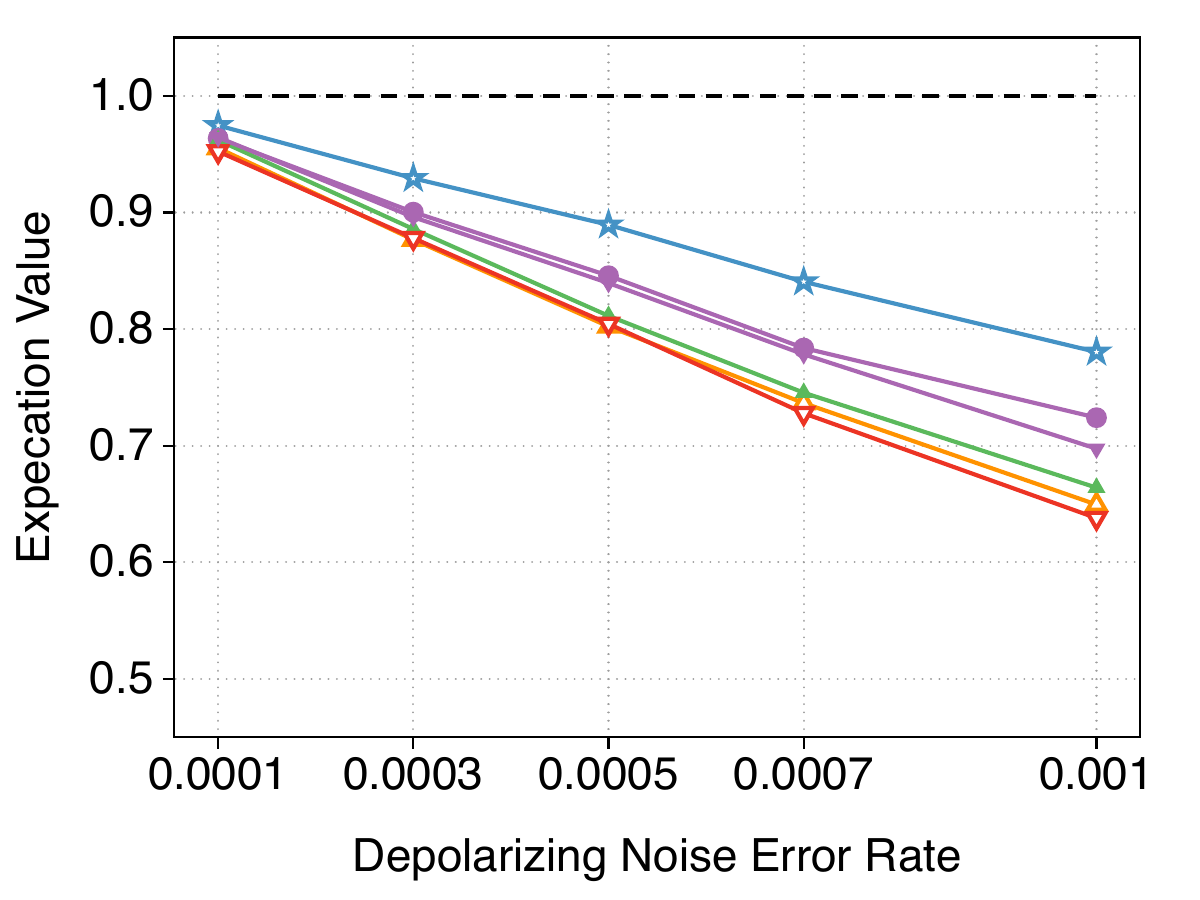}
        \label{fig:bv_n14}
    \end{subfigure}
    \begin{subfigure}[t]{0.32\columnwidth}
        \centering
        \caption{bv\_n19' Circuit}
        \includegraphics[width=\columnwidth]{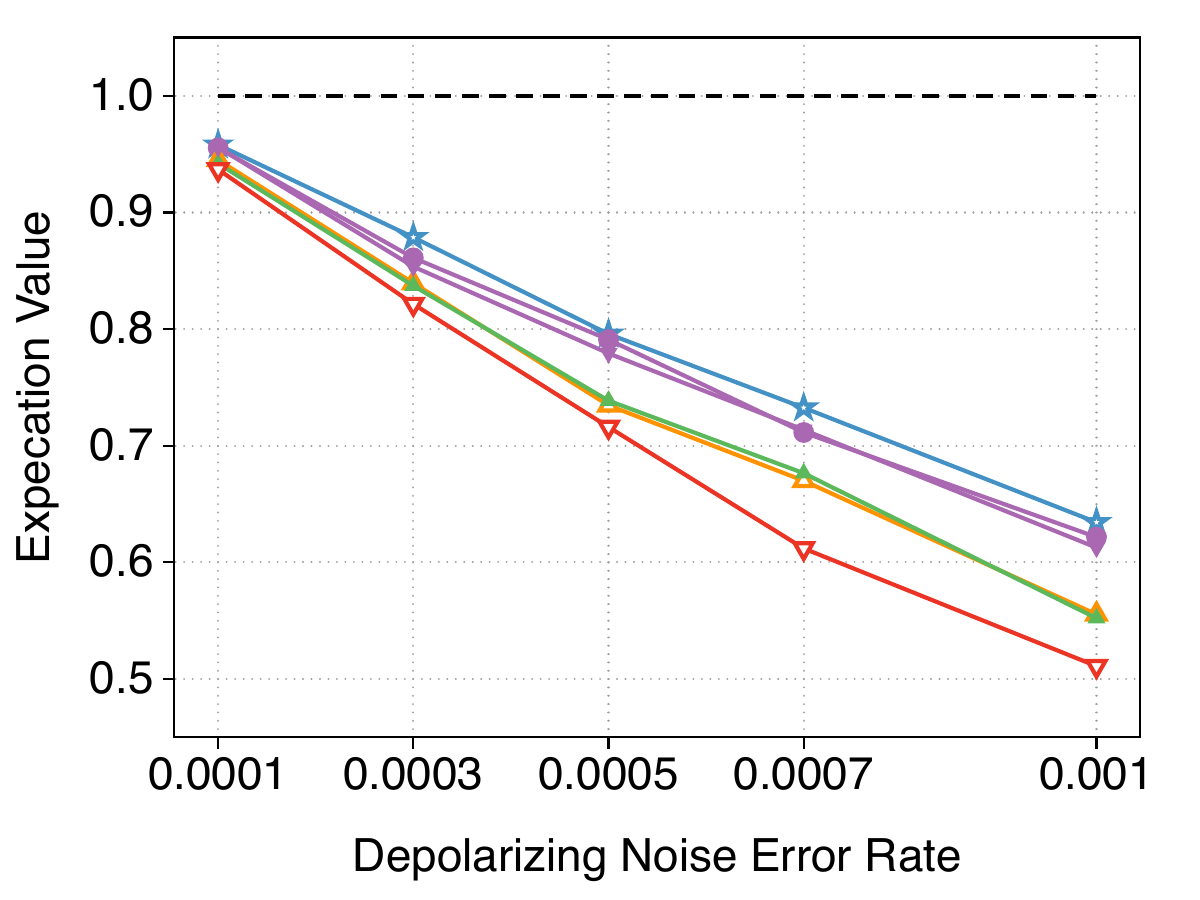}
        \label{fig:bv_n19}
    \end{subfigure}
    \begin{subfigure}[t]{0.32\columnwidth}
        \centering
        \caption{multiplier\_n15' Circuit}
        \includegraphics[width=\columnwidth]{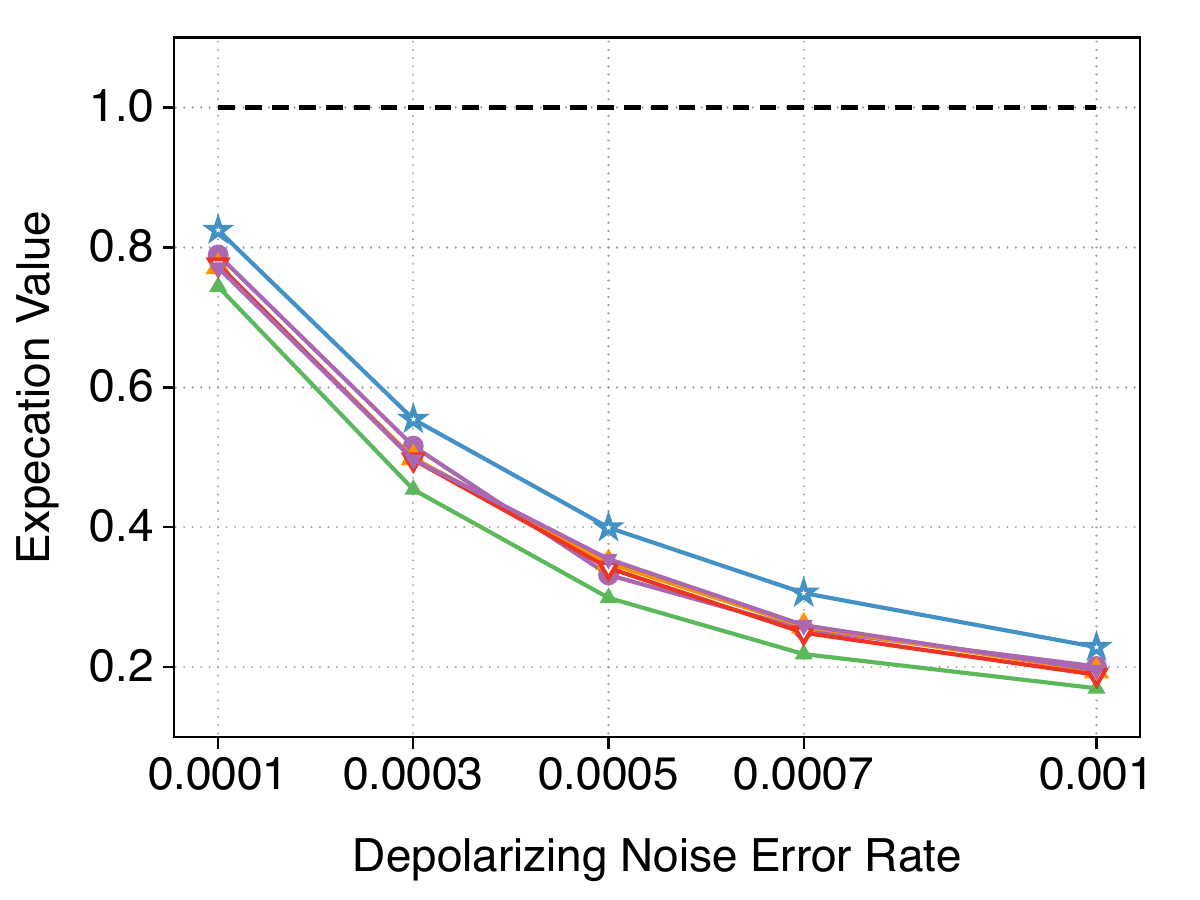}
        \label{fig:multipler_n15}
    \end{subfigure}
    \begin{subfigure}[t]{0.32\columnwidth}
        \centering
        \caption{multiply\_n13' Circuit}
        \includegraphics[width=\columnwidth]{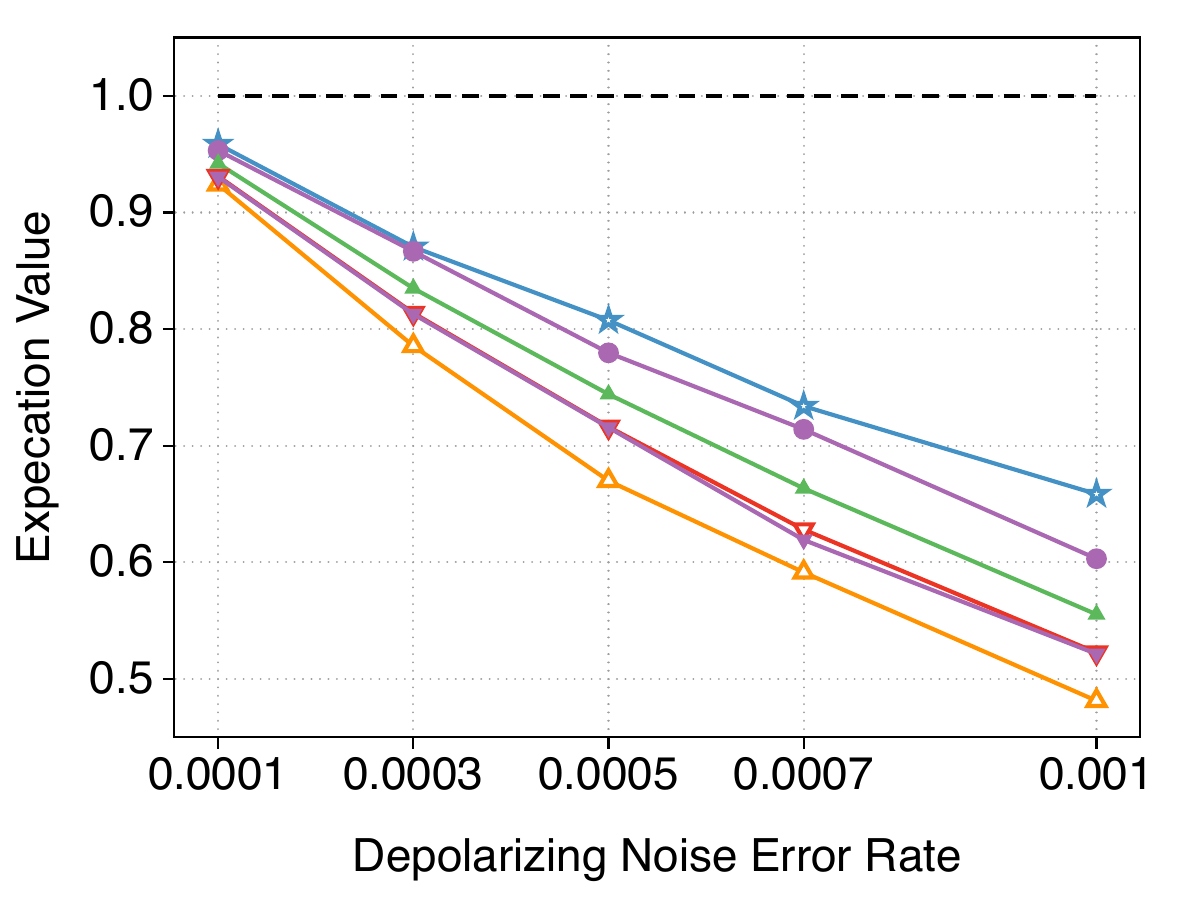}
        \label{fig:qram_n20}
    \end{subfigure}
    \caption{Experiments involve applying a depolarizing noise model to execute the NISQ benchmark quantum circuit on CA-CORE. These results are then compared with outcomes obtained from various topologies. The collected results are compared with the Exact Simulation's expectation value, which remains unaffected by noise. Consequently, the fidelity of exact simulation is 1. A result from various topologies closer to 1 indicates a higher fidelity.}
    \label{fig:noisy_benchmark}
\end{figure*}
\subsection{Fidelity Improvement}\label{sec:eval_noise}
To assess noise impact on CA-CORE, we use a depolarizing noise model with an error rate of $\epsilon$. 
This model assumes uniform error rates for all gates: $\epsilon$ for one-qubit channels and $\epsilon \times 5$ for two-qubit channels like the CNOT gate, making the latter five times more error-prone.

Figure~\ref{fig:noisy_benchmark} depicts the performance of our CA-CORE architecture compared with those of various topologies mentioned in Section~\ref{sec:experiment_setup}. 
For IBM topologies, we use the original configuration, however, for Google's Sycamore, which has 53 qubits, efficient simulation using a noise model is not feasible. Therefore, we employ a half-Sycamore topology setup with six rows and four columns, with all couplers enabled. 

CA-CORE outperforms existing topologies in all evaluations. Particularly noteworthy is the 40.13\% improvement in fidelity at an error rate ($\epsilon$) of 0.001 compared with the result achieved by Sycamore, as shown in Figure~\ref{fig:qec9xz}. 
On average for all error rates, CA-CORE exhibits fidelity improvements of 15.76\%, 16.39\%, 7.52\%, 5.69\%, 13.91\%, and 9.77\%, as shown in Figure~\ref{fig:bigadder}, ~\ref{fig:qec9xz}, ~\ref{fig:bv_n14}, ~\ref{fig:bv_n19},  ~\ref{fig:multipler_n15}, and~\ref{fig:qram_n20}, respectively.

Notably, in the result shown in Figure~\ref{fig:bv_n19} for the `bv\_n19' circuit, the improvement is not significant at lower error rates. 
The `bv\_n19' circuit represents the Bernstein-Vazirani algorithm \cite{bernstein}.
This algorithm requires a brief quantum circuit with two H gates for all qubits, one NOT (X) gate, and $n-1$ CNOT gates, where $n$ is the qubit count.
Each CNOT gate connects all qubits with the last one. This structure ensures that at lower error rates, errors across different setups have minimal impact on result fidelity.
Further, Google Sycamore (24) performs worse than all the other topologies. 
This is primarily because the experiment employed a trivial qubit-mapping method. 
For the compiled circuit on various topologies, the circuit depths are as follows: 40 for CA-CORE, 43 for ibmq\_almaden, 49 for ibmq\_johannesburg, 48 for ibmq\_cairo, 53 for Google's Sycamore (24), and 45 for ibmq\_prague. 
With differences in circuit depth, the impact of relatively high error rates also increases, contributing to the inferior performance of Google Sycamore's topology. 
This contributes to the marginal improvement seen in CA-CORE compared to existing topologies across low-to-higher error-rate configurations.

\subsection{Scalable Time Complexity}\label{sec:eval_execution_time}
In the reprocessing step, an input graph of $n$ qubits is prioritized, where each qubit can connect to any other. 
This phase identifies all qubit correlations and their coupling weights with a time complexity of $O(n^2)$. 
Then, Algorithm~\ref{alg:max_weighted_path_gen} constructs a new graph to maximize the weighted path using the obtained qubit connections from the preprocessing phase.

In the new path graph, each qubit aligns to have a maximum of two adjacent connections, excluding loop connections.
Next, the path graph is converted into a structured grid graph using Algorithm~\ref{alg:grid_topology_generation}, specifying the desired columns and rows, with a time complexity of $O(n)$ for this transformation.
The final step involves connecting the grid's qubits based on the discovered connections during preprocessing, with a time complexity of $O(E)$, where $E$ represents the original graph's connections.
Furthermore, constraint removal adds another $O(E)$ in the worst case to calculate the weight of the violated constraint on the connectivity. Therefore, the total time complexity of the algorithm is $O(n^2 + n + 2E)$. 
With this time complexity, our algorithm can efficiently reconfigure the topology for the input quantum circuit.  
It takes less than 1$s$ to configure a topology for the 33-qubit quantum circuit used in benchmark Table~\ref{tb:SWAP_reduction}.
\begin{figure}
    \centering
    \begin{subfigure}[t]{0.25\columnwidth}
        \centering
        \includegraphics[width=\columnwidth]{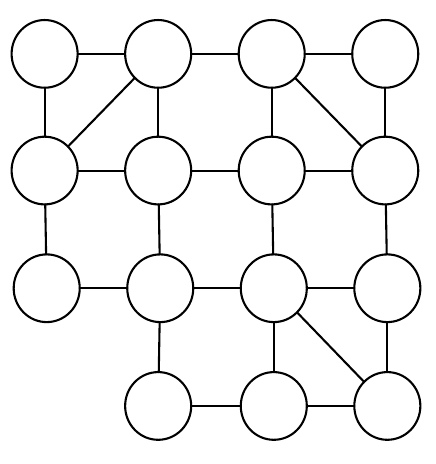}
        \caption{`multiplier\_n15'}
        \label{fig:multipler_n15_topology}
    \end{subfigure}
    \hspace{20pt}
    \begin{subfigure}[t]{0.35\columnwidth}
        \centering
        \includegraphics[width=\columnwidth]{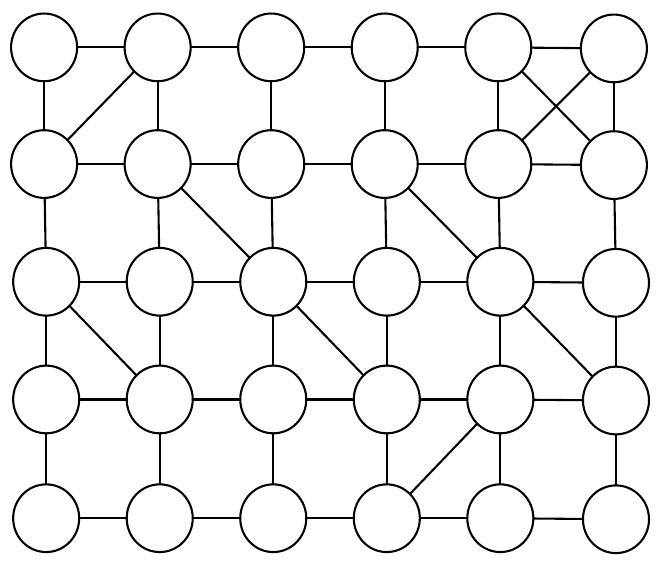}
        \caption{`Random Circuit 30-Qubit'}
        \label{fig:random_circuit_30_topology}
    \end{subfigure}
    \caption{Our topology reconfigured for specific quantum circuits. (a) Reconfigured topology for `multipler\_n15' circuit benchmark in Figure \ref{fig:multipler_n15}. (b) Reconfigured topology for the randomly generated 30-qubit circuit benchmark in Table \ref{tb:SWAP_reduction}. }
    \label{fig:reconfigured_topology}
\end{figure}

\section{Discussion and Future Work}\label{sec:discussion}
This study explores methods to adjust qubit connections within a tunable coupler SQC implemented in a grid topology setup. The adoption of a grid-like structure is influenced by previous studies \cite{YANG2023106944, li2020te} demonstrating high yield with optimal connectivity.
Specifically, we formalize the configured topology in three steps, each of which considers the correlations between qubits. 
To the best of our knowledge, this is the first study to utilize a quantum circuit for efficient coupling configuration while addressing the frequency-collision constraint of the SQC.

The manuscript refrains from directly comparing the proposed qubit mapping methodology with existing approaches due to a fundamental divergence in their focal points. 
Existing qubit mapping strategies predominantly concentrate on fixed topology target quantum hardware, aiming to optimize within the constraints of predetermined qubit layouts. 
In contrast, the manuscript emphasizes configuring optimal qubit coupling maps, specifically targeting tunable coupler-based quantum hardware. 
As a result, the comparison isn't directly pursued as the existing qubit mapping techniques are tailored to a different paradigm, focusing on fixed hardware configurations, whereas the manuscript's methodologies are designed for systems with tunable coupler features, warranting a distinct evaluation approach

While our proposed method efficiently identifies an optimal coupling map, reducing the need for additional SWAP operations and enhancing the fidelity of the input quantum circuit, further improvements and experiments are required.

\textbf{Integrating Qubit Mapping Optimization}: Many qubit mapping problems consider fixed connectivity on the target hardware. 
While our evaluation employs a basic qubit mapping approach, we anticipate that a well-designed qubit mapping solution, which considers the reconfigurable topology of the target hardware, can yield a significantly optimized topology.

\textbf{Design Space and Frequency Allocation}: Our study adheres to the design space, constraints, and frequency allocation configuration outlined in references \cite{YANG2023106944, li2020te}. 
As quantum hardware fabrication techniques advance, additional studies should be undertaken to accommodate diverse quantum hardware configurations.


\textbf{Real QC Evaluation}: Our study outlines innovative methods for optimizing qubit coupling maps in quantum circuits, emphasizing efficiency and error reduction. 
While simulations form a strong foundation, exploring disparities between simulated and real-world outcomes would enrich the manuscript. 
Analyzing trade-offs between runtime coupler tuning and swap gate introduction offers insights for developers and users. Assessing how optimized layouts impact specific quantum algorithms clarifies practical implications. 
Looking ahead, our pioneering work in configuring optimal physical qubit coupling map layouts requires experimental validation on real quantum hardware. 
Collaborations with quantum computing experts or accessing tunable-coupler devices provide invaluable insights for real quantum computing. 
Investigating scalability, refining hardware-based constraints, and exploring error mitigation strategies are essential for future research, bridging theory with practical applications.

\section{Related Work}\label{sec:related_work}

Various studies have been conducted on tunable coupler-based SQCs and qubit coupling map layouts~\cite{rob2022mc,li2020te, Leroux2021}.
Tunable two-qubit couplers have shown potential for application in mitigating errors within multi-qubit gate in SQC processors~\cite{Leroux2021}. 
Additionally, architectures have been proposed for both the modular implementation of tunable couplers~\cite{Campbell2023} and the utilization of floating tunable couplers in designs of large-scale quantum processors~\cite{Sete2021}.

These studies primarily emphasize the design of tunable couplers to enhance the scalability of SQC processors.
Gushu et al.~\cite{li2020te} emphasized a logical qubit coupling topology and constructed the hardware architecture using three subroutines: layout design, bus selection, and frequency allocation. 
This was achieved while considering physical constraints based on the profiling information from the coupling degree list and the coupling strength matrix. 
Their primary objective was to enhance the yield rate by maximizing qubit connectivity. 
However, they did not address the noise implications of their architecture.
Tan et al.~\cite{Tan2022qm} presented QLSA-reconfigurable-atomic-arrays (RAA), a layout synthesizer RAA during the QC compilation phase. 
While their method achieves gate reduction and heightened fidelity for reconfigurable architectures through hardware-adaptive constraints, it is specifically designed for atomic arrays. 
Conversely, our proposed approach addresses the constraints inherent in SQCs.

Lin et al.~\cite{lin2022ds} proposed a method similar to ours which utilizes layout synthesis to establish qubit edge connections. 
However, the edges that they determined contradict the fabrication constraints due to the frequency collision outlined in \cite{YANG2023106944, li2020te}. 
Conversely, our proposed method considers the edge connection that adheres to the frequency collision constraint, maximizing the SQC's connectivity. 
Given these contradictions and disparities in architectural design for edge selection, we cannot conduct a comparative study between their method and our method.

\section{Conclusion}\label{sec:conclusion}
Quantum algorithms demand extensive resources that often surpass the available quantum hardware capacity.
However, the design of application-specific architectures for quantum algorithms can help reduce the required quantum resources while enhancing result fidelity. 
This study delves into the benefits of initializing an optimal layout for a target quantum algorithm on a tunable coupler-based SQC to eliminate unnecessary connections between uncorrelated qubits. 
This involves an analysis of the input circuit to determine the contextual correlation matrix between qubits and their connectivity weights. 
Subsequently, we propose an algorithm to configure a grid-like topology, considering qubit correlations and the physical constraints of the SQC. 
Experimental results demonstrate that the context-aware reconfigured topology can reduce circuit depth by minimizing additional SWAP operations in a randomly generated quantum circuit with a complex pattern while improving output fidelity in practical NISQ benchmark circuits.

\begin{ack}
    This work was supported by Institute for Information \& Communications Technology Planning \& Evaluation (IITP) grant funded by the Korea government (MSIT) (No. 2020-0-00014, A Technology Development of Quantum OS for Fault-tolerant Logical Qubit Computing Environment).
\end{ack}

\section*{Reference}
\bibliographystyle{iopart-num}
\bibliography{manuscript}

\end{document}